\documentstyle[multicol,prb,aps,psfig]{revtex}

\newcommand{\beq}{\begin{equation}}
\newcommand{\eeq}{\end{equation}}
\newcommand{\eqna}{\begin{eqnarray}}
\newcommand{\eqne}{\end{eqnarray}}
\newcommand{\eqnaa}{\begin{eqnarray*}}
\newcommand{\eqnae}{\end{eqnarray*}}
\newcommand{\dia}{\begin{displaymath}}
\newcommand{\die}{\end{displaymath}}

\begin{document}
\title{{\bf {\ Statistics of Hartree-Fock Levels in Small Disordered Systems}}}
\author{Shimon Levit$^{1}$ and Dror Orgad$^{2}$}
\address{$^{1}$ Department of Condensed Matter Physics, The Weizmann Institute of 
Science,\\
Rehovot 76100, Israel\\
$^{2}$ Department of Physics and Astronomy, University of California 
Los Angeles,\\
Los Angeles, CA, 90095-1547\\}
\date{\today}

\maketitle

\begin{abstract}
We study the statistics of quasiparticle and quasihole levels in small
interacting disordered systems within the Hartree-Fock approximation. The
distribution of the inverse compressibility, given according to Koopmans'
theorem by the distance between the two levels across the Fermi
energy, evolves from a Wigner distribution in the non-interacting limit to a
shifted Gaussian for strong interactions. On the other hand the nature of
the distribution of spacings between neighboring levels on the same side of
the Fermi energy (corresponding to energy differences between excited states
of the system with one missing or one extra electron) is not affected by the
interaction and follows Wigner-Dyson statistics. These results are derived
analytically by isolating and solving the appropriate Hartree-Fock equations
for the two levels. They are substantiated by numerical simulations of the
full set of Hartree-Fock equations for a disordered quantum dot with Coulomb
interactions. We find enhanced fluctuations of the inverse compressibility
compared to the prediction of Random Matrix Theory, possibly due to
localization of the wavefunctions around the edge of the dot. The
distribution of the inverse compressibility calculated from the discrete
second derivative with respect to the number of particles of the
Hartree-Fock ground state energy deviates from the distribution of the level
spacing across the Fermi energy. The two distributions have similar shapes
but are shifted with respect to each other. The deviation increases with the
strength of the interaction thus indicating the breakdown of Koopmans'
theorem in the strongly interacting limit.
\end{abstract}

\pacs{PACS numbers: 73.20.Dx, 05.45+b, 31.15.N}

\begin{multicols}{2}

\section{Introduction.}

It is universally accepted that the principal signature of quantum chaos is
the statistics of Random Matrix Theory (RMT), Refs. \cite{Meh,Wei} which is
obeyed by the energy levels of chaotic systems. This is supported by
semiclassical considerations \cite{Ber} as well as many numerical \cite
{Ber,Boh} and analytical \cite{Efe} examples. However realistic chaotic
systems such as quantum dots, small metallic grains or the so called yrast
levels in rotating nuclei involve interactions between many particles
(electrons, nucleons, etc). Interesting problem then arises of how the
interactions are expected to modify the RMT predictions. Several recent
experimental and theoretical publications began to deal with this problem 
\cite{Siv,Sim,Marcus,Pep,Imr,Gef,Bla,BerAlt,Koul,ber1,Kam,Mat}.

Applications of RMT to noninteracting chaotic systems, such as quantum dots,
are concerned with statistical properties of single particle quantities.
Analogous and experimentally relevant in interacting systems are quantities
which characterize quasiparticles. In small disordered systems one can
discuss the statistics of their energy levels, lifetimes and wave functions
(respectively real and imaginary parts and the residues of the poles of the
single particle Green's function). Experience gained in nuclear and atomic
physics indicates that the Hartree-Fock (HF) method provides a very
reasonable approximate description of quasiparticle properties in finite
systems. The non degenerate HF particle-hole excitations form a convenient
basis to describe low lying excitations of these systems. In this paper we
adopt this description and study statistical properties of the HF levels in
small disordered systems. Properties of charged excitations have been probed
experimentally by measuring Coulomb blockade in disordered quantum dots, cf.
Refs. \cite{Siv,Sim,Marcus}. The neutral particle-hole excitations can be
measured by studying acoustic phonon \cite{Yehoshua} and microwave absorption 
\cite{Alt+Spivak}.

\section{The Hartree-Fock Approximation in Weakly Disordered Systems.}

Interacting electrons in a disordered system are described by the
Hamiltonian 
\begin{equation}
H=\sum_{\alpha\beta}h_{\alpha\beta}a_\alpha^{\dagger}a_{\beta} + \frac{1}{2}%
\sum_{\alpha\beta\gamma\delta}V_{\alpha\beta\gamma\delta}
a_{\alpha}^{\dagger}a_{\beta}^{\dagger}a_{\delta}a_{\gamma}\; ,  \label{mnyh}
\end{equation}
where $|\alpha\rangle$, $|\beta\rangle, \dots $ denote states of a single
particle basis. The non interacting part of $H$ is controlled by the one
body Hamiltonian $h_{\alpha\beta}$ representing the disordered system. In
this work we are interested in the regime of disorder for which 
the random matrix $h_{\alpha\beta}$ can be
viewed as described by the rules of RMT. The interaction $%
V_{\alpha\beta\gamma \delta}$ is not random and in a quantum dot, for
example, represents matrix elements of the Coulomb or screened Coulomb
interaction. In our discussion however we will regard it as a general matrix
and will be able to draw conclusions for wide classes of possible $%
V_{\alpha\beta\gamma\delta}$.

Our main approximation will be to treat the Hamiltonian (\ref{mnyh}) in the
HF approximation. The HF equations are 
\begin{equation}
\sum_{\beta }h_{\alpha \beta }^{(HF)}\psi _{i}(\beta )=\varepsilon _{i}\psi
_{i}(\alpha )\;,  \label{oldeq}
\end{equation}
with 
\begin{equation}
h_{\alpha \beta }^{(HF)}=h_{\alpha \beta }+\sum_{\alpha ^{\prime }\beta
^{\prime }}V_{\alpha \alpha ^{\prime }\beta \beta ^{\prime }}^{A}\rho
_{\beta ^{\prime }\alpha ^{\prime }}\;,
\end{equation}
or symbolically $h^{(HF)}=h+tr(V^{A}\rho )$. Here $V^{A}$ is the
antisymmetrized interaction matrix $V_{\alpha \beta \gamma \delta
}^{A}\equiv V_{\alpha \beta \gamma \delta }-V_{\alpha \beta \delta \gamma }$
and the trace is taken over the second pair of the indices of $V^{A}$. 
The self-consistent density matrix $\rho$ is given by 
\begin{equation}
\rho _{\alpha \beta }=\sum\limits_{h\in holes}\psi _{h}(\alpha )\psi
_{h}^{*}(\beta )\, .  \label{oldro}
\end{equation}
Henceforth we
will denote by $h$, $h^{\prime }$ and $p$, $p^{\prime }$, etc., the hole 
(occupied) and particle (unoccupied) levels respectively.

It is common to view the HF approximation as arising from the variational
minimization of the total energy of the system. In this interpretation the
direct physical meaning of the energies $\varepsilon _{i}$ and the wave
functions $\psi _{i}(\alpha )$ remains obscure and one must use the so
called Koopmans theorem, Ref. \cite{Koopmans}. We recall however that the HF
equations can also be derived from an approximation to the equation of
motion for the one particle Green's function 
\end{multicols}
\widetext

\noindent
\setlength{\unitlength}{1in}
\begin{picture}(3.375,0)
  \put(0,0){\line(1,0){3.375}}
  \put(3.375,0){\line(0,1){0.08}}
\end{picture}
\begin{eqnarray}
G(\alpha ,\beta ;\omega ) &=&-i\int_{-\infty }^{\infty }dt\ e^{i\omega
t}\langle \Phi _{0}(N)|Ta_{\alpha }(t)\ a_{\beta }^{+}(0)|\Phi
_{0}(N)\rangle \ =  \nonumber \\
&=&-\sum_{i}\frac{\langle \Phi _{0}(N)|a_{\alpha }|\Phi _{i}(N+1)\rangle
\langle \Phi _{i}(N+1)|a_{\beta }^{+}|\Phi _{0}(N)\rangle }{%
E_{i}(N+1)-E_{0}(N)-\omega -i0} \\
&&-\sum_{i}\frac{\langle \Phi _{0}(N)|a_{\beta }^{+}|\Phi _{i}(N-1)\rangle
\langle \Phi _{i}(N-1)|a_{\alpha }|\Phi _{0}(N)\rangle }{E_{0}(N)-E_{i}(N-1)
-\omega +i0} \; .  \nonumber
\end{eqnarray}
The HF energies $\varepsilon _{p\text{ }}$and $\varepsilon _{h}$ and the
corresponding wave functions $\psi _{p}(\alpha )$ and $\psi _{h}(\alpha )$
are then approximate energies and wave functions of respectively
quasiparticles and quasiholes, 
\begin{eqnarray}
\varepsilon _{p\text{ }} \simeq E_{p}(N+1)-E_{0}(N) \;\;\; &;& \;\;\;
\varepsilon_{h}\simeq E_{0}(N)-E_{h}(N-1)  \nonumber \\
\psi _{p}(\alpha ) \simeq \langle\Phi_0(N)|a_{\alpha}|\Phi_p(N+1) \rangle
\;\;\; &;& \;\;\; \psi_{h}(\alpha )\simeq\langle \Phi_h(N-1) |a_{\alpha
}|\Phi_0(N)\rangle \; .  \nonumber
\end{eqnarray}
\hfill
\begin{picture}(3.375,0)
  \put(0,0){\line(1,0){3.375}}
  \put(0,0){\line(0,-1){0.08}}
\end{picture}

\begin{multicols}{2}
\noindent

We wish to study the statistical properties of the set $\varepsilon _{i}$
and the wave functions $\psi _{i}(\alpha )$ which follow from the random
nature of $h_{\alpha \beta }$. Ideally one would like to be able, starting
from the probability distribution $P(h)$, to determine the joint probability
distribution $P(\varepsilon _{1},\varepsilon _{2},...,\varepsilon _{n},...)$
and similar distribution for $\psi _{i}(\beta )$ and on its basis to predict
various correlation properties of these quantities. In the present paper we
address a much simpler problem of the repulsion pattern of neighboring pairs
of $\varepsilon _{i}$ and its application to the addition spectra of quantum
dots. We will present simple analytic approximations and perform numerical
investigations to check their validity.

\section{Theoretical Considerations.}

\subsection{The Constant Part of the Interaction.}

The simplest limiting case is a constant interaction $V(|\mbox{\bf r}-
\mbox{\bf r}^{\prime }|)=V_{0}$ which will serve as a reference point for
our discussion. In the following we consider the spinless case. One has for
the matrix elements in (\ref{mnyh}) 
\begin{equation}
V_{\alpha \beta \gamma \delta }=V_{0}\delta _{\alpha \gamma }\delta _{\beta
\delta }  \label{cnsin}
\end{equation}
This interaction allows for a trivial exact solution which is reproduced by
the HF equations. They are solved by the eigenfunctions $\psi
_{i}^{(0)}(\alpha )$ of the one body Hamiltonian 
\begin{equation}
h_{\alpha \beta }\psi _{i}^{(0)}(\beta )=\varepsilon _{i}^{(0)}\psi
_{i}^{(0)}(\alpha )\;,
\end{equation}
and have the following eigenvalues 
\begin{equation}
\varepsilon _{i}=\left\{ 
\begin{array}{r@{\quad : \quad}l}
\varepsilon _{i}^{(0)}+V_{0}(N-1) & n_i=1 \\ 
\varepsilon _{i}^{(0)}+V_{0}N & n_i=0 
\end{array}
\right. \;,
\end{equation}
where $N$ is the number of particles and $n_i$ is the occupation number 
of the i'th level. It is clear that the lowest energy 
solution is obtained by occupying the lowest $N$ non-interacting states.
The gap which separates the energies
of the occupied (hole) and empty (particle) levels for $V_0>0$ 
results from the absence
of the contribution from the exchange term in the latter. In a sense one can
say that the empty levels do not create the exchange hole which decreases
(increases) the level energy for positive (negative) $V_{0}$.

From the known distribution of $\varepsilon _{i}^{(0)}$'s it is easy to
calculate the statistical properties of the HF levels $\varepsilon _{i}$.
Since the spacings of neighboring levels are $s=\Delta \varepsilon ^{(0)}$
for levels below and above $\varepsilon _{f}$ their distribution is given by
the ordinary Wigner surmise (e.g. $P(s)=P_{W}(s)\propto s\exp [-\pi (\frac{s
}{2<s>})^{2}]$ for GOE). The spacing between the levels lying across the
Fermi energy is $s=V_{0}+\Delta \varepsilon ^{(0)}$ and therefore its
distribution is given by the Wigner surmise shifted by the value $V_{0}$, 
\begin{equation}
P(s)=P_{W}(s-V_{0})\;.  \label{sftwg}
\end{equation}
The gap in the distribution of the level spacings across the Fermi energy is
reflected in the differences of the total ground state energy $E_{0}$ of the
system as a function of the particle number 
\begin{eqnarray}
&&E_{0}(N) =\sum_{i=1}^{N}\varepsilon _{i}^{(0)}+\frac{V_{0}}{2}N(N-1)\;,
\label{gsenergy} \\
&&E_{0}(N+1)-E_{0}(N) =\varepsilon _{N+1}^{(0)}+V_{0}\,N=\varepsilon_{N+1}
\;, \label{gsediff}\\
&&D_{2}(N) \equiv E_{0}(N+1)-2E_{0}(N)+E_{0}(N-1) \nonumber \\
&&\hspace{1.15cm} =\varepsilon
_{N+1}^{(0)}-\varepsilon _{N}^{(0)}+V_{0}=\varepsilon_{N+1}-\varepsilon_{N}
\label{D2def}\;.
\end{eqnarray}
Here we defined the quantity $D_{2}(N)$ which is essentially the inverse
compressibility of the dot. Its distribution has been measured in transport
experiment of quantum dots in the Coulomb blockade regime, Refs. \cite
{Siv,Sim,Marcus}. Eq. (\ref{gsediff}) is Koopmans theorem which is exact 
in this case. It leads directly to the expression (\ref{D2def}) of $D_{2}$ 
in terms of the energy difference between the two HF levels across the Fermi 
energy, a fact which  motivates the study of this spacing. 
The Coulomb gap in $D_{2}$ does not fluctuate in the limit
of constant interaction. For more realistic interactions the Coulomb
blockade gap must undergo fluctuations in addition to the fluctuations of
the single particle energies $\varepsilon _{i}$. Study of these fluctuations
will be one of our principal goals.

\subsection{Realistic Interaction -- Repulsion of the Hartree-Fock Levels.}

For small $s$ ($s\ll \Delta $ -- the mean level spacing) the well known
behavior of the distribution $P(s)$ for random Hamiltonians (i.e. linear for
GOE, quadratic for GUE, etc.) can be derived by considering the repulsion of
close pairs of levels. In this subsection we generalize this analysis to the
case of two close HF levels. The new element in the HF problem, Eq. (\ref
{oldeq}), is the presence of the non linear self-consistent term 
$\sum_{\alpha ^{\prime }\beta ^{\prime }}V_{\alpha \alpha^{\prime }\beta
\beta ^{\prime }}^{A}\rho_{\beta ^{\prime }\alpha ^{\prime }}$ which
implicitly depends on the realization of the random part $h_{\alpha\beta }.$

Let us assume that for some realization of $h_{\alpha \beta }$ the HF
Hamiltonian $h_{\alpha \beta }^{(HF)}$ has two closely lying levels, 
$\varepsilon _{2}>\varepsilon _{1}$, so that $\Delta \varepsilon \equiv
\varepsilon _{2}-\varepsilon _{1}\ll \Delta $. We wish to investigate how 
$\Delta \varepsilon $ reacts when the random $h$ is varied by $\delta h$ with 
$\Delta \gg |\delta h_{IJ}|\sim\Delta \varepsilon $ for $I,J=1,2$.

The variation of $h$ causes a change $\delta h^{(HF)}=\delta
h+tr(V^{A}\delta \rho )$ in the HF Hamiltonian where the second term is due
to the induced change of the self-consistent density. Without this term one
would get for the new spacing the standard result $s=2\sqrt{(\Delta
\varepsilon +\delta h_{22}-\delta h_{11})^{2}+4|\delta h_{12}|^{2}}\;,$
which for future reference we shall rewrite in the form 
\begin{equation}
s=2\sqrt{B_{1}^{2}+B_{2}^{2}+B_{3}^{2}}\equiv 2|\mbox{\bf B}|\;.
\label{sumsqr}
\end{equation}
The vector $\mbox{\bf B}$ is a projection on the Pauli matrices of the 
$2\times 2$ Hamiltonian $(h+\delta h)_{IJ}$ in the subspace of the two close
levels, $I,J=1,2$, 
\begin{equation}
\mbox{\bf B}=\frac{1}{2}tr[\mbox {\boldmath $\sigma$}(h+\delta
h)]\;\;\;,\;\;\;\sigma _{i}-{\rm the\;\;Pauli\;\;matrices}\;.
\end{equation}
The level repulsion at small spacings is a consequence of the
proportionality of $s$ to the square root of the sum of squares of real
quantities and can be deduced by evaluating, for small $s$ 
\begin{equation}
P(s)=\int \delta (s-s({\bf B}))P({\bf B})d{\bf B}\;.  \label{psint}
\end{equation}
The familiar behavior $P(s)\sim s$ for GOE and $P(s)\sim s^{2}$ for GUE 
$(s\ll \Delta )$ follows then directly for any $P({\bf B})$ which does not
vanish for small $B=|{\bf B}|$. The (unjustified) evaluation of (\ref{psint}) 
for all values of $s$ taking $P(B)$ to be of a Gaussian form results in
the Wigner distribution for the case of real $h$.

In the Appendix we discuss how the result (\ref{sumsqr}) is modified by the
presence of the self-consistent term $tr(V^{A}\delta \rho )$. Under the
conditions discussed in Section \ref{valid} we show there that :

1. When the two close levels are both occupied or both empty the
distribution of their level spacings for small $s\ll \Delta$ is $\sim s$ or 
$\sim s^2$ as in the non interacting case.

2. When the two levels are on opposite sides of the Fermi level, i.e. one
occupied and one empty the expression (\ref{sumsqr}) changes to 
\begin{equation}
s=2\left| \mbox{\bf B}-\stackrel{\leftrightarrow }{\bf J}\cdot \ \mbox{\bf m}
(\mbox{\bf B},\stackrel{\leftrightarrow }{\bf J})\right| \;.  \label{lvspint}
\end{equation}
Here we use the dyadic notation $[\stackrel{\leftrightarrow }{\bf J}\cdot 
{\bf m}]_{a}=\sum_{b=1}^{3}J_{ab}m_{b}$. As in the non-interacting case the
``vector'' ${\bf B}$ contains the information about the random part 
$h+\delta h$ of the HF Hamiltonian whereas the matrix $\stackrel
{\leftrightarrow }{\bf J}$ is the projection on the Pauli matrices of the
interaction matrix $\widetilde{V}_{IJKL},\;(I,J,K,L=1,2)$ in the subspace of
the two close levels $\psi _{1}$ and $\psi _{2}$ (cf. Appendix for the
precise definition of $\widetilde{V}$, ${\bf B}$ and $\stackrel
{\leftrightarrow }{\bf J}$). The vector $\mbox{\bf m}(\mbox{\bf B}, \stackrel
{\leftrightarrow }{\bf J})$ is a unit vector $|\mbox{\bf m}|=1$ and is a
solution of the self-consistent HF equations for the two levels 
\begin{equation}
{\bf m}\times ({\bf B}-\stackrel{\leftrightarrow }{\bf J}\cdot \ {\bf m}
)=0\;.  \label{vfse}
\end{equation}

In this notation the total HF energy is

\begin{equation}
{\cal E}={\bf B\cdot m}-\frac{1}{2}{\bf m}\cdot \stackrel{\leftrightarrow }
{\bf J}\cdot {\bf m}\;.  \label{tehf}
\end{equation}

The full investigation of the solutions of the above equations and the
expression (\ref{lvspint}) is found in the Appendix. Here we will
concentrate on the simplest case when the matrix of interactions $J_{ab}$ is
degenerate, i.e. has equal eigenvalues. As is shown in the Appendix such a
degenerate $J_{ab}$ matrix corresponds to a complete absence of degeneracy
(such as spin) of the close HF levels. We will furthermore
restrict ourselves to real Hamiltonians $h+\delta h$. Then the vector 
$\mbox{\bf B}$ has only two components $B_{1}$ and $B_{3}$ and one needs to
consider only the corresponding components of $J_{ab}$ with $a,b=1,3$. For a
degenerate $J_{ab}$ the vector $\stackrel{\leftrightarrow }{\bf J}\cdot {\bf 
m}$ is parallel to ${\bf m}$ and the two level HF equation (\ref{vfse})
becomes a simple linear relation $m_{1}B_{3}-m_{3}B_{1}=0$ without any
dependence on the interaction. The level spacing (\ref{lvspint}) however
still contains the interaction term. Using the normalization condition 
$m_{1}^{2}+m_{3}^{2}=1$ one obtains 
\begin{equation}
s=2|B+J|\;,  \label{gaps}
\end{equation}
where $B=\sqrt{B_{1}^{2}+B_{3}^{2}}$ and $J$ is the degenerate eigenvalue.
The explicit expression for $J$ is 
\begin{equation}
J=(V_{1212}-V_{1221})/2\;.  \label{exprJ}
\end{equation}

For a positive $J$ which corresponds to the case of a repulsive interaction
the equation for the spacing (\ref{gaps}) describes a conical
surface separated by a distance $2J$ from the $(B_1,B_3)$ plane. This
means that the probability for a given $s$ has a gap of the magnitude $2J$
above which it must rise linearly 
\begin{equation}
P(s;J)=\left\{ 
\begin{array}{cc}
0\hspace{2cm}\;\; & s<2J \\ 
(s-2J)f(s)\;\; & s\ge 2J
\end{array}
\right. \;,  \label{pshf}
\end{equation}
where $f(s)$ is an undefined function which is finite at $s=2J$. For
general reasons one should expect that $f(s)$ vanishes at $s\to \infty $.

The result (\ref{pshf}) implies that the distribution $P(s;J)$ is
qualitatively the same as the shifted Wigner distribution 
\begin{equation}  \label{shiftwig}
P(s;J)= \left\{ 
\begin{array}{cc}
0\hspace{3.95cm} & s<2J \\ 
\frac{\pi}{2\Delta^2}(s-2J)e^{-\frac{\pi}{4}\left(\frac{s-2J}{\Delta}
\right)^2} \;\; & s\ge 2J
\end{array}
\right. \; ,
\end{equation}
similar to what was found in the case of the constant interaction, Eq. (\ref
{sftwg}). There is however a crucial difference which we will now try to
elucidate.

As is seen from Eq. (\ref{exprJ}) the degenerate eigenvalue $J$ is invariant
under unitary transformations in the subspace spanned by the two given close
states $\psi _{1}$ and $\psi _{2}$. As long as one examines realizations of 
the random $h$ for which these two particular states stay close in energy 
they mix strongly only between themselves and $J$ stays the same. The
distribution of the level spacings for such selection of $h$'s is given by (%
\ref{pshf}) with the fixed $J$ as in the constant interaction case. However
variations of $h$ may bring another pair of levels, say, $\psi _{1}$ and $%
\psi _{3}$ at close distance on both sides of $\varepsilon _{f}$. For such
levels the value of $J$ may be completely different and given by the
corresponding matrix elements $(V_{1313}-V_{1331})/2$. In calculations of
the overall $P(s)$ for small $s$ one should therefore average over the
distribution $P(J)$ of all $J$'s 
\begin{equation}
P(s)=\int P(J)P(s;J)dJ\;.  \label{foldJ}
\end{equation}

For a constant interaction $J=V_{0}$ for any pair of states $\psi _{i}$, $%
\psi _{j}$ and therefore $P(J)$ is a delta function centered at $V_{0}$. For
an extreme short range interaction $V(|\mbox{\bf r}-\mbox{\bf r}^{\prime
}|)=V_{0}\delta (\mbox{\bf r}-\mbox{\bf r}^{\prime })$ all values of $J$ are
zero and do not fluctuate. One expects therefore that for a general
interaction the average value of $J$ is coming mainly from the very long
range component while its fluctuations reflect the middle range components
of $V$.

The main feature of matrix elements (\ref{cnsin}) of a constant interaction
is that they have the same form in any (orthonormal) basis of the single
particle states ${\psi _{\alpha }}$ which are used to calculate $V_{\alpha
\beta \gamma \delta }$. In other words for a constant interaction the matrix
elements $V_{\alpha \beta \gamma \delta }$ are {\em invariant under the
unitary group U(M) of all transformations in the entire single particle
Hilbert space of the problem}. Here $M$ is the dimensionality of this space
(typically $M\to \infty $). This invariance of $V_{\alpha \beta \gamma
\delta }$ for a constant interaction is the reason that $P(J)$ is a delta
function in this case. In order to develop a theory of $P(J)$ for a general
interaction one must understand the statistical properties of the single
particle wave functions--solutions of the HF problem. The matrix elements in
(\ref{exprJ}) are defined with respect to such wave functions. Although we
do not have an analytic theory of $P(J)$ we will present below numerical
investigation of this function and the validity of the expression of the
type (\ref{foldJ}). Numerically we find that the distribution of $J$ is
approximately Gaussian. Adopting this finding into expression (\ref{foldJ})
and assuming the form (\ref{shiftwig}) for $P(s;J)$ one obtains 
\end{multicols}
\widetext

\noindent
\setlength{\unitlength}{1in}
\begin{picture}(3.375,0)
  \put(0,0){\line(1,0){3.375}}
  \put(3.375,0){\line(0,1){0.08}}
\end{picture}
\begin{equation}  \label{gaussps}
P(s)=\beta\sqrt{\frac{\alpha}{\alpha+\beta}}\, e^{-\frac{\alpha\beta} 
{\alpha+\beta}\left(s-2J_0\right)^2} 
\times\left\{\frac{\alpha}{\alpha+\beta}\, {\rm erfc}\left[-\frac{\alpha}
{\sqrt{\alpha+\beta}}\left(s-2J_0\right)\right] +\frac{1}{\sqrt{%
\pi(\alpha+\beta)}}\, e^{-\frac{\alpha^2}{\alpha+\beta}\left(
s-2J_0\right)^2}\right\} \; ,
\end{equation}
\hfill
\begin{picture}(3.375,0)
  \put(0,0){\line(1,0){3.375}}
  \put(0,0){\line(0,-1){0.08}}
\end{picture}

\begin{multicols}{2}
\noindent
where $\alpha=1/8{\sigma_J}^2$ and $\beta=\pi/4\overline\Delta^2$ with $%
\overline\Delta$ being the mean level spacing at the vicinity of the Fermi
energy. $J_0$ and $\sigma_J$ are the mean and standard deviation of $P(J)$
respectively.

\subsection{Validity of the Two Level Treatment -- Why Only the Statistics
of Spacings Across the Fermi Level is Effected by the Interaction.\label%
{valid}}

The above results are valid as long as the spacing between the given
two levels is much smaller than the distance to other levels. This
restriction is needed in order to be able to isolate the two levels from the
rest. For repulsive interactions the distance between the pair of levels on 
different sides of the Fermi energy is determined by the size of the matrix 
element $J$ (cf.,Eq. \ref{gaps}). Therefore the condition of the validity of 
our treatment for such levels is $J\ll \Delta$.
As we will see below numerical evidence indicates that at least
qualitatively the expression (\ref{foldJ}) remains correct also for a much
larger $J\sim \Delta $. For an even stronger interaction the two level
treatment ceases to be valid and one must account for the reaction of
distant levels to the changes of the self-consistent potential 
$tr(V^{A}\delta \rho ).$

Here we wish to add the following remark. For a constant interaction our
result is trivially valid for any strength. On the other hand for any given
interaction one can extract a constant part, i.e. $V_{\alpha \beta \gamma
\delta }=V_{0}\delta _{\alpha \gamma }\delta _{\beta \delta }+U_{\alpha
\beta \gamma \delta }$, where $U_{\alpha \beta \gamma \delta }=V_{\alpha
\beta \gamma \delta }-V_{0}\delta _{\alpha \gamma }\delta _{\beta \delta }$.
Since the HF wavefunctions are independent of $V_0$ one may try to solve the 
problem using $U$ first and then add $V_0$. This procedure is not unambiguous 
and must be dictated by the physics of the problem. If $P(J)$ vanishes for 
$J<J_c$, i.e. in the presence of a hard gap 
(like in a quantum dot with Coulomb interaction), it is natural to take 
$V_0=J_c$. The condition of validity is then $J_r\ll \Delta$ where $J_r$ 
is the typical gap due to the residual interaction $U$.

When there is no minimal value for $J$ any subtraction is bound to produce 
$U$ which has both attractive and repulsive components. This fact may 
introduce fundamental differences between the solutions to the HF
problem obtained using the interaction $V$ and the one derived in the manner 
indicated above. Most notably using the latter procedure one will find 
cases for which there exists enhanced probability for the two levels across 
$\varepsilon_f$ to be close 
to each other [or at distance $V_0$ after the addition of the constant part, 
see result (\ref{psjatt}) of the appendix]. However if the width $\sigma_J$ of 
$P(J)$ is much smaller then its mean $J_0$ the use of $V_0=J_0-\sigma_J$ will 
cause $U$ to change sign only in a small number of cases. Consequently we 
may use in such a case $\sigma_J\ll \Delta$ as our criterion for the validity 
of the two level treatment.   

For pairs of neighboring levels which lie on the same side of the Fermi
energy the change in the density matrix $\delta\rho$ 
vanishes as long as only the two level subspace is considered. Consequently 
there is no self-consistent term $V^{A}\delta \rho$ present and one 
recovers the results of the non-interacting case.  
There are two types of corrections to this two level treatment
which must be considered. One is the non-zero contribution of the distant
levels to $\delta \rho .$ Another is the correction to the energies of
the two close levels due to virtual transition to distant levels.

To estimate the first correction we use the first order result
\begin{equation}
\delta \rho _{\alpha \beta }=\sum_{ph}\left[ \frac{\langle p|\delta
h+V^{A}\delta \rho |h\rangle }{\varepsilon _{h}-\varepsilon _{p}}\psi
_{p}(\alpha )\psi _{h}^{*}(\beta )+h.c.\right] \;,  \label{delrho1st}
\end{equation}
which can be solved to obtain the RPA expression
\end{multicols}
\widetext

\noindent
\setlength{\unitlength}{1in}
\begin{picture}(3.375,0)
  \put(0,0){\line(1,0){3.375}}
  \put(3.375,0){\line(0,1){0.08}}
\end{picture}
\begin{equation}
\delta \rho _{\alpha \beta }=\sum_{\alpha ^{^{\prime }}\beta ^{^{\prime
}}}\left[ 1-\sum\limits_{p^{^{\prime }}h^{^{\prime }}}\left( \frac
{\langle p^{^{\prime }}|V^{A}|h^{^{\prime }}\rangle}{\varepsilon_
{h^{^{\prime}}}-\varepsilon _{p^{^{\prime }}}}\psi_{p^{^{\prime }}}\psi
_{h^{^{\prime}}}^{*}+h.c.\right) \right] _{\alpha \beta ,\alpha^{^{\prime }}
\beta^{^{\prime }}}^{-1}  
\times\sum\limits_{ph}\left[ \frac{\langle p|\delta h|h\rangle}
{\varepsilon _{h}-\varepsilon _{p}}\psi _{p}(\alpha ^{\prime })\psi
_{h}^{*}(\beta ^{\prime })+h.c.\right] \; .  \label{drho2}
\end{equation}
\hfill
\begin{picture}(3.375,0)
  \put(0,0){\line(1,0){3.375}}
  \put(0,0){\line(0,-1){0.08}}
\end{picture}

\begin{multicols}{2}
\noindent
Thus we find
\begin{equation}
\delta \rho \sim \frac{\delta h}{\Delta+V} \; ,
\label{esti1}
\end{equation}
where here $V\sim\sum_{p',h'}V_{pp'hh'}$.
The contribution to $\delta h^{(HF)}$ is therefore not just $\delta h$ but 
$\delta h^{(HF)}\sim \delta h[1+V/(V+\Delta )].$ However this still means 
that for such levels $\delta h^{(HF)}\sim \delta h$ for any strength of 
interaction $V$, the sole role of which is to renormalize the random part 
$\delta h$.

The corrections to the energies of the close levels due to transitions to 
distant levels are
\begin{equation}
E_{I}=\varepsilon _{I}+\sum_{i\neq 1,2}\frac{|\delta h_{Ii}^{(HF)}|^{2}}{%
\varepsilon _{I}-\varepsilon _{i}}\ ,\hspace{0in}\;\;\ I=1,2  \label{tlcor}
\end{equation}
where the $\varepsilon _{I}$'s are the energies obtained in the two level
treatment. But we have just shown that $\delta
h^{(HF)}\sim \delta h$. Therefore the correction term in expression 
(\ref{tlcor}) is of order $(\delta h)^{2}/\Delta \ll \delta h$
and can be disregarded for any $V$. Thus we expect that the two level
treatment for a couple of occupied or empty levels 
gives correctly the small $s$ behavior of $P(s)$ for any interaction
strength. To stress, the difference between this case and the case of two 
levels on both sides of $\varepsilon _{f}$ is that there $\delta \rho $ also
included a ''non perturbative'' zero order term coming from the mixing of
the wave functions of the occupied and the empty states.

\subsection{Addition Spectrum vs Excitation Spectrum.}

The HF energies are interpreted as energies of quasiparticles (for $%
\varepsilon _{i}>\varepsilon _{f}$) or quasiholes ($\varepsilon
_{i}<\varepsilon _{f}$). They are excitations of a system with one added or
one subtracted particle, i.e. $E_{i}(N\pm 1)-E_{0}(N)=\pm \varepsilon _{i}$.
The excitations with the same particle number $E_{i}(N)-E_{0}(N)$ are
described in the HF approximation by solutions representing determinants
with the same particle number $N$ which are orthogonal to the HF ground state
determinant. In the case of a constant interaction such excitations are
simply particle-hole excitations of the non-interacting problem. The
distribution of their spacings does not show any Fermi energy related gap
and coincides with the non shifted Wigner distribution. This is also true in
the two level HF model, Eq. (\ref{vfse}). Let us demonstrate this in the
simple case of the degenerate $\stackrel{\leftrightarrow }{\bf J}$. The HF
equation (\ref{vfse}) in this case is $m_{1}B_{3}-m_{3}B_{1}=0$ which
together with the normalization condition $m_{1}^{2}+m_{3}^{2}=1$ produce
two solutions with different total HF energies, Eq. (\ref{tehf}), ${\cal E}%
=\pm B-J/2$. These solutions represent the ground and excited states of this
model which have the same number of particle. The linear dependence of the
difference $\Delta {\cal E}=2B$ indicates that at least as long as the two
level treatment of the level repulsion is valid the HF energy spacing
distribution between such states obeys $P(\Delta {\cal E})\sim \Delta {\cal E%
}$ for small $\Delta {\cal E}$ without any gap, as in the non-interacting
systems. 

Yet another way to obtain this result is to consider the spacings
between neighboring particle or hole levels. They correspond, within the HF
approximation, to the distances between adjacent excited states of the
system with $N+1$ and $N-1$ electrons respectively. By the arguments given
in the preceding subsection these spacings follow the Wigner-Dyson statistics.

\subsection{A Schematic Model -- Keeping Only the Average Interaction Matrix
Elements.}

For a given realization of the random $h_{\alpha \beta }$ let us consider
the Hamiltonian in the eigenbasis of $h$, for which  $h_{\alpha \beta
}=\varepsilon _{\alpha }^{(0)}\delta _{\alpha \beta }$. In this random basis
also the matrix elements of the interaction $V_{\alpha \beta \gamma \delta }$
are random. Their statistical properties are known, cf., Refs.\cite{Ranint}
and are as follows. Only the matrix elements $V_{\alpha \beta \alpha \beta }$
and $V_{\alpha \beta \beta \alpha }$ have non zero averages. Their
distributions are narrow with the width behaving like $1/M$ in the Random
Matrix Theory (M - the size of the single particle space) and like $1/g$ in
the random potential theory (g - dimensionless conductance). Based on
these properties one is tempted to approximate the interaction by retaining
only the matrix elements with non zero averages, i.e. to assume that 
\begin{equation}
V_{\alpha \beta \gamma \delta }=\delta _{\alpha \gamma }\delta _{\beta
\delta }\ V_{1}^{(\alpha \beta )}+\delta _{\alpha \delta }\delta _{\beta
\gamma }\ V_{2}^{(\alpha \beta )} \; ,
\end{equation}
with $V_{1}^{(\alpha \alpha )}=V_{2}^{(\alpha \alpha )}$. Such a model has
an easy exact solution. The Hamiltonian 
\begin{equation}
H=\sum_{\alpha }\varepsilon _{\alpha }^{(0)}\,\hat{n}_{\alpha }+\frac{1}{2}%
\sum_{\alpha, \beta}J_{\alpha \beta }\,\hat{n}_{\alpha }\,%
\hat{n}_{\beta }  \label{sch0} \; ,
\end{equation}
where $J_{\alpha \beta }=V_{1}^{(\alpha \beta )}-V_{2}^{(\alpha \beta )},$
has exact eigenstates given by the eigenfunctions of the occupation
operators $\hat{n}_{\alpha }$ 
\begin{equation}
\psi =|n_{1},n_{2},\cdots ,n_{k},\cdots\rangle \; , 
\end{equation}
with the corresponding eigenenergies 
\begin{equation}
E(\{n_{i}\})=\sum_{i}\varepsilon _{i}^{(0)}\,n_{i}+\frac{1}{2}%
\sum_{i,j}J_{ij}\,n_{i}\,n_{j} \; .
\end{equation}
Like in the case of the constant interaction the exact results are reproduced 
by the HF approximation. The HF equations are 
\begin{equation}
\left[\varepsilon _{\alpha }^{(0)}+\sum_{\beta}J_{\alpha \beta }\,
 \rho _{\beta \beta }\right]\! \phi_{i}(\alpha )-\sum_{\beta}J_{\alpha \beta }
\, \rho _{\alpha \beta }\,  \phi _{i}(\beta)=\varepsilon _{i}\phi _{i}
(\alpha )  \label{hfeqsh}
\end{equation}
and are solved by $\phi _{i}(\alpha )=\delta _{i\alpha }.$ Therefore 
\begin{equation}
\varepsilon _{i}=\varepsilon _{i}^{(0)}+\sum_{h}J_{ih}\; ,
\label{hfenerschem}
\end{equation}
where the sum is over the occupied levels. 

We are interested in the ground state of the Hamiltonian (\ref{sch0}). 
A general argument due to Ref.\cite{Lieb} guarantees that at least for 
positive definite (repulsive) interactions, which we assume below, 
the HF ground state must be comprised of the $N$ lowest energy single 
particle HF levels $\phi_1,\cdots,\phi_N$. While the ground 
state of the constant interaction model is obtained by filling the $N$ 
lowest non-interacting states this need not be the case for the present 
model. Consider
\begin{eqnarray}
\nonumber
\varepsilon _{N+1}-\varepsilon _{N}=&&\varepsilon _{N+1}^{(0)}-\varepsilon
_{N}^{(0)}+J_{N+1,N} \\
&&+\sum_{k=1}^{N-1}(J_{N+1,k}-J_{N,k})\; .  \label{sch1}
\end{eqnarray}
Although $\varepsilon _{N+1}^{(0)}-
\varepsilon_{N}^{(0)}+J_{N+1,N}$ is positive definite the sum in (\ref{sch1})
is over $2N-2$ random variables and can be large and negative. Consequently 
it may happen that $\varepsilon _{N+1}<\varepsilon _{N}$ in variance with the 
above condition on the ground state. In such a case a different occupation 
pattern must be sought (we note that even if the condition is not violated 
there may exist other solutions that are consistent with it and which have 
lower energies). We expect such crossings of levels across the Fermi energy 
to take place when $\Delta+J_0\simeq\sqrt{2N}\sigma_J$, where $J_0$ and 
$\sigma_J$ are the mean and standard deviation of $J_{\alpha\beta}$ 
respectively.

Similar crossings may occur for levels below or above $\varepsilon _{f}$ 
when $\Delta\simeq\sqrt{2N}\sigma_J$ since for them  
\begin{equation}
\varepsilon _{j}-\varepsilon _{j-1}=\varepsilon _{j}^{\left( 0\right)
}-\varepsilon _{j-1}^{\left( 0\right) }+\sum_{k=1}^{N}(J_{jk}-J_{j-1,k}) \; . 
\label{sch2}
\end{equation}
This shuffling of levels tends to reduce the correlation between the energies 
of neighboring states. In particular one expects to find weaker level 
repulsion when either $N$ or the interaction strength ($\sigma_J$) 
are increased. 

\begin{figure}
  \narrowtext  
\setlength{\unitlength}{1in}
\begin{picture}(3.2,4.7)(0,-1)  
   \put(-0.25,-0.8){\psfig{figure=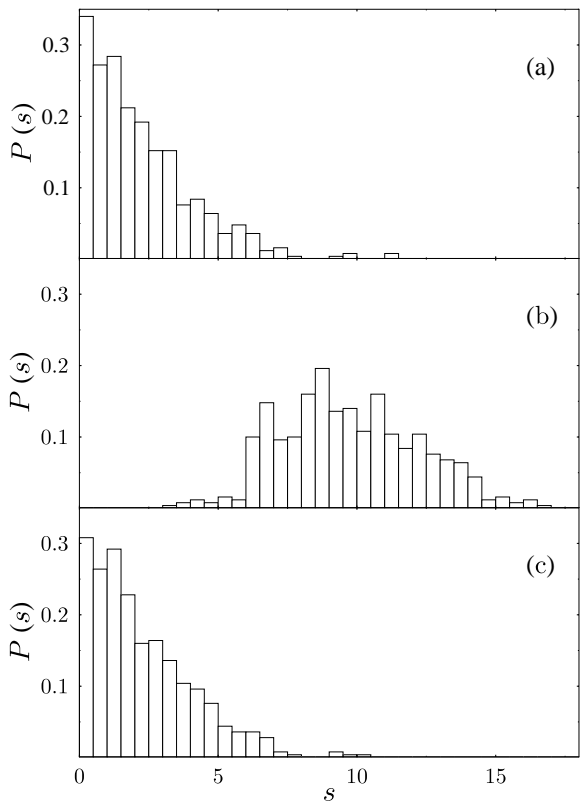,width=3.75in}}
\end{picture}
\caption{Probability density distributions of the spacings between a) the 
last two hole levels below $\varepsilon_{f}$, b) the two levels across 
$\varepsilon_{f}$ and c) the first two particle levels above 
$\varepsilon_{f}$. The results were derived by solving the HF equations of 
the schematic model Eq.({\protect\ref{sch0}}). $\varepsilon_{\alpha}^{(0)}$ 
and $J_{\alpha\beta}$ were generated by the random potential model with 
Coulomb interaction, described in Section {\protect\ref{numericalres}}, 
for a dot with 15 electrons, $W/t=1.2$ and $U/t=1.2$. 
We also included a constant interaction part of strength $V_0=6\Delta_0$.
The spacings are measured in terms of the non-interacting mean level 
spacing $\Delta_0$.}
\label{schemfig}
\end{figure}

To verify the above discussion we have calculated the HF energies 
(\ref{hfenerschem}) using numerical values for $\varepsilon _{\alpha }^{(0)}$ 
and $J_{\alpha \beta }$ generated by the random potential model described in 
the next Section. We took care to choose the lowest energy solution. The 
results for $P(s)$ below, across and above $\varepsilon_f$ are shown in 
Fig. \ref{schemfig}. These distributions differ significantly from the 
exact HF distributions calculated while retaining the off-diagonal elements
of $V$, cf., Figs. \ref{debgraph},\ref{deagraph},\ref{deefgraph}.
The most prominent incorrect feature is the absence of level repulsion in 
$P(s)$ for the levels below or above $\varepsilon _{f}$. One can attempt to
correct this feature by using the average rather than the exact matrix
elements which enter $J_{\alpha \beta}.$ As it follows from the random
potential model, Refs.\cite{Ranint}, such average matrix elements are
functions of the corresponding eigenenergies, i.e. $J_{\alpha \beta
}=j(|\varepsilon _{\alpha }-\varepsilon _{\beta }|)$ with $j(x)$ a known
function which is approximately $ -\ln x$ in two dimensions. Although this
model reproduces the density of states in the vicinity of the Fermi energy 
it is not expected to account for the correct correlations between neighboring 
levels as manifested in $P(s)$.

\section{Comparison with Numerical Results.}
\label{numericalres}

In order to substantiate the results of our analytical two level treatment
we have numerically solved the complete set of HF equations (\ref{oldeq})
derived from a tight-binding Hamiltonian for a disordered two dimensional
quantum dot. With the labeling of the sites by a double index $(i,j)$ the
one-body Hamiltonian is given by 
\begin{equation}
h=\! \sum_{i,j}\varepsilon _{i,j}a_{i,j}^{\dagger }a_{i,j}-t\!\sum_{i,j}\left[
a_{i+1,j}^{\dagger }a_{i,j}+a_{i,j+1}^{\dagger }a_{i,j}+h.c.\right] \, ,
\label{tbh}
\end{equation}
where $\varepsilon _{i,j}$ is the energy of the site $(i,j)$ and $t$ is a
constant hopping matrix element. Each of the energies $\varepsilon _{i,j}$
is chosen randomly from a Gaussian distribution with the standard deviation $%
W/2$. We assume repulsive $1/r$ interaction 
\begin{equation}
V=U\sum_{i,j,k,l}{\frac{a_{i,j}^{\dagger }a_{k,l}^{\dagger }a_{k,l}a_{i,j}}{|%
{\bf r}_{i,j}-{\bf r}_{k,l}|/b}}\;,  \label{tbv}
\end{equation}
where $b$ is the lattice constant and $U=e^{2}/b$.

The dot was approximated by a grid of $M=20\times 20=400$ sites with hard
wall boundary conditions. Most of the numerical data was obtained for a dot
filled with $N=15-17$ spinless electrons and a disorder strength $W=1.2\,t$.
Under such conditions the dot was in the diffusive regime and the levels in
the vicinity of $\varepsilon _{f}$ exhibited RMT statistics in the
non-interacting limit. For the low filling that we used the energy band was
approximately parabolic and the Fermi energy for a clean non-interacting
system was $\varepsilon _{f}=4\pi tN/M$. A convenient dimensionless measure
of the strength of the interaction is $r_{s}=(e^{2}/a)/\varepsilon _{f}$
where $a=\sqrt{M/\pi N}b$ is the average inter-particle distance. In our
case $r_{s}=(U/t)\sqrt{M/16\pi N}\simeq 0.7\,U/t$. Below we describe our
results for $U/t=0.2-1.6$. Henceforth energy is quoted in units of the
observed non-interacting mean level spacing at the Fermi energy $\Delta _{0}$
which was found to be larger by 7\% then the clean value $4\pi t/M$.

\begin{figure}
  \narrowtext  
\setlength{\unitlength}{1in} 
\begin{picture}(3.25,2.5)(0,-1)
\put(-0.4,-1.2){\psfig{figure=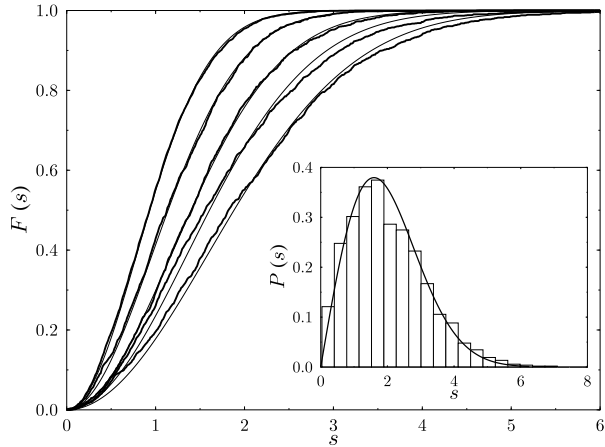,width=4in}}
\end{picture}      
\caption{Probability distributions $F(s)=\int_0^s P(s^{\prime})\,
ds^{\prime} $ of spacings between the last two hole levels just below $
\varepsilon_{f}$. The heavy lines depict from top to bottom the numerical
results for $U/t=0, 0.4, 0.8, 1.2$ and 1.6. The solid curves present the
best fit to the data assuming a Wigner function with renormalized mean level
spacing $\overline {\Delta}$. The inset contains $P(s)$ for $U/t=1.6$
(histogram) together with its best fit to a Wigner function.}
\label{debgraph}
\end{figure}

\begin{figure}
  \narrowtext 
\setlength{\unitlength}{1in}
\begin{picture}(3.25,2.5)(0,-1)  
\put(-0.4,-1.2){\psfig{figure=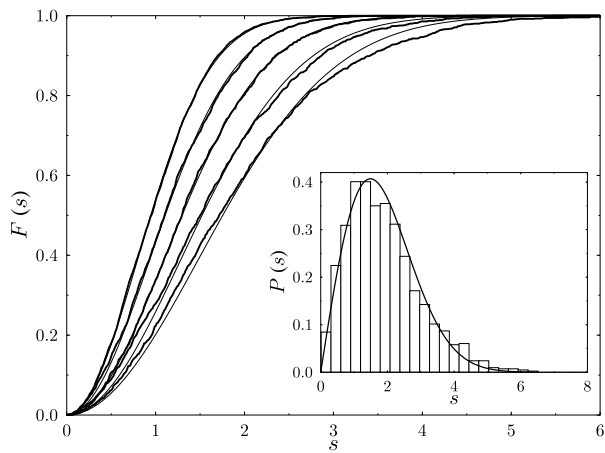,width=4in}} 
\end{picture}     
\caption{Same as Fig.{\protect\ref{debgraph}} but
for the spacing between the first two particle 
levels just above the Fermi energy.}
\label{deagraph}
\end{figure}

\begin{figure}
  \narrowtext 
\setlength{\unitlength}{1in}
\begin{picture}(3.25,2.5)(0,-1)  
\put(-0.4,-1.2){\psfig{figure=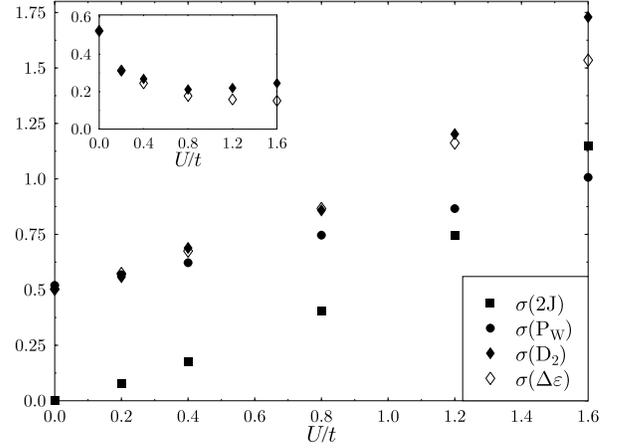,width=4in}} 
\end{picture}     
\caption{The standard deviations of $2J$, $P_W$ - the Wigner distribution that
fits best the distribution of spacings in the vicinity of $\epsilon_f$, $%
D_2(16)$ and $\Delta\epsilon$ - the spacing between the two levels across
the Fermi energy. The inset depicts the normalized fluctuations $%
\sigma(D_2)/ \langle D_2\rangle$ (full diamonds) and $\sigma(\Delta%
\epsilon)/\langle \Delta\epsilon\rangle$ (empty diamonds). The results are
for a dot with 15-17 electrons and disorder strength $W=1.2\, t$.}
\label{vargraph}
\end{figure}

The distribution functions $F(s)=\int_0^s P(s^{\prime})\, ds^{\prime}$ of
the spacings between the last two occupied (hole) levels and between the
first two empty (particle) levels are shown in Figs. \ref{debgraph} and \ref
{deagraph}. They are compared with the best fit to an integrated Wigner
function with a renormalized mean level spacing $\overline{\Delta }$. These
and the following results were obtained by averaging the HF spectrum of
15,16 and 17 electrons over 450-500 realizations of the disorder. While the
distributions vanish quadratically for small spacings increasing deviations
from the Wigner function are observed when the interaction becomes stronger.
We find enhanced probability for the occurrence of spacings smaller and much
larger then $\overline{\Delta }$ for large values of $U/t$. The renormalized
mean level spacing $\overline{\Delta }$ also increases with the strength of
the interaction. The width of the fitted Wigner function $\sigma(P_{W})=0.52\,
\overline{\Delta }$ is presented in Fig. \ref{vargraph}. It grows 
approximately linearly with $U/t$. The mean level spacing
between adjacent levels further away from the Fermi energy decreases with
the distance from $\epsilon_f$ and approaches 1.

\begin{figure}
  \narrowtext 
 \setlength{\unitlength}{1in}
\begin{picture}(3.25,2.75)(0,-1)  
    \put(-0.05,-0.9){\psfig{figure=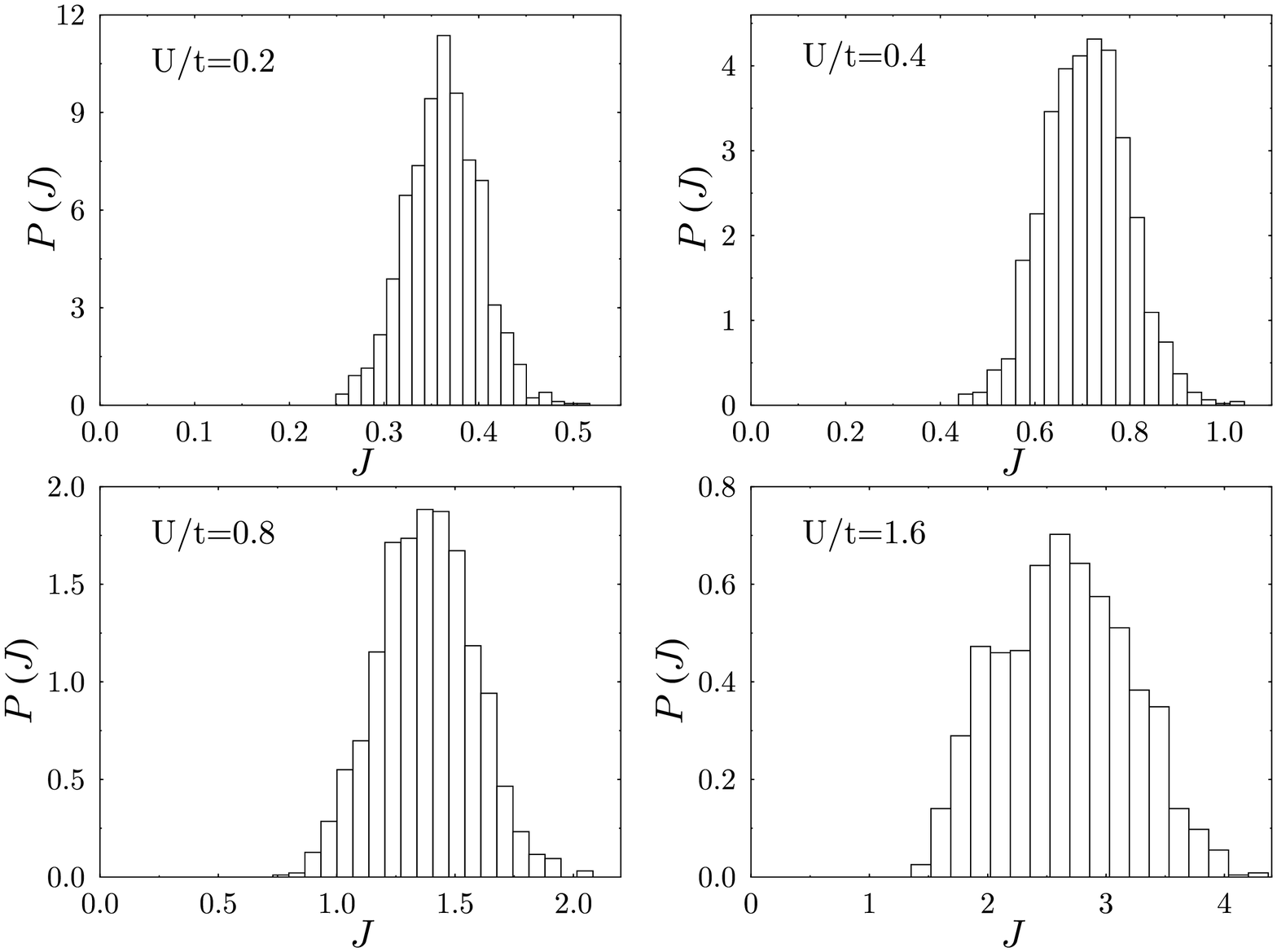,width=3.32in}}
\end{picture}      
\caption{Probability density distributions of the parameter $J$ for
various interaction strengths.}
\label{jgraph}
\end{figure}

\begin{figure}
  \narrowtext 
\setlength{\unitlength}{1in}
\begin{picture}(3.25,2.5)(0,-1)  
\put(-0.4,-1.2){\psfig{figure=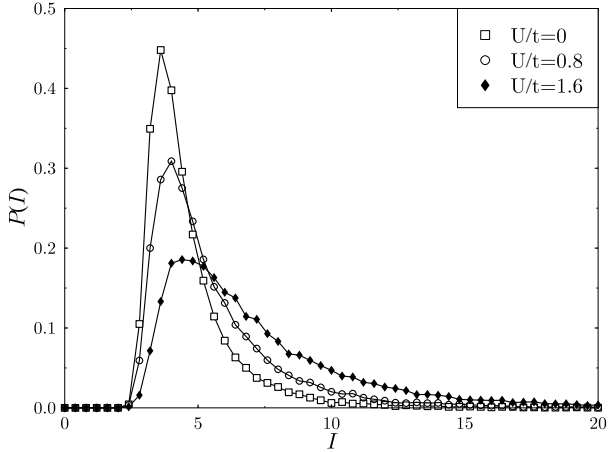,width=4in}} 
\end{picture}     
\caption{Probability density distributions of the inverse participation
ratio of the 10 hole and 10 particle levels around $\varepsilon_f$ 
for various interaction strengths.}
\label{iprgraph}
\end{figure}

The distributions of the quantity $J$ defined in Eq. (\ref{exprJ}) are
depicted in Fig. \ref{jgraph}. They are approximately Gaussian for small
values of the interaction strength but develop asymmetry towards the high $J$
end when $U/t$ is increased. The mean of the distribution scales with the
interaction as $\langle J\rangle \simeq 1.7\,U/t$. For the width $\sigma (J)=%
\sqrt{\langle J^{2}\rangle -\langle J\rangle ^{2}}$ we find $\sigma
(J)\simeq 0.16\,U/t+0.13\,(U/t) ^{2}$ over the range of parameters studied
(see Fig. \ref{vargraph}). These fluctuations are responsible for the
smearing of the distribution of spacings across $\varepsilon_f$ as will be
shown below. A lengthy calculation using the random vector model (RVM) gives
for our system $\langle J\rangle=2.1\,U/t$ and $\sigma (J)=0.032\,U/t$. The
fact that the RVM result for $\langle J\rangle $ is larger then the observed
one reflects the tendency of the system to prefer a non-uniform density
distribution that reduces the Coulomb energy. We believe that this is also
the reason for at least part of the enhancement of the actual fluctuations 
relative to the RVM predictions. Typically we found the HF eigenfunctions 
(both particles and holes) to have large amplitude along the periphery of the 
dot, as expected from simple electrostatic considerations. The localizing 
effect of the interaction is evident from Fig. \ref{iprgraph} where we 
present the distribution of the inverse participation ratio $I=Mb^{4}
\sum_{i,j}\psi^{4}(i,j)$ averaged over the 10 particle and 10 hole levels 
around $\epsilon_f$.

\begin{figure}
  \narrowtext 
\setlength{\unitlength}{1in}
\begin{picture}(3.2,4.8)(0,-1)  
\put(-0.25,-0.8){\psfig{figure=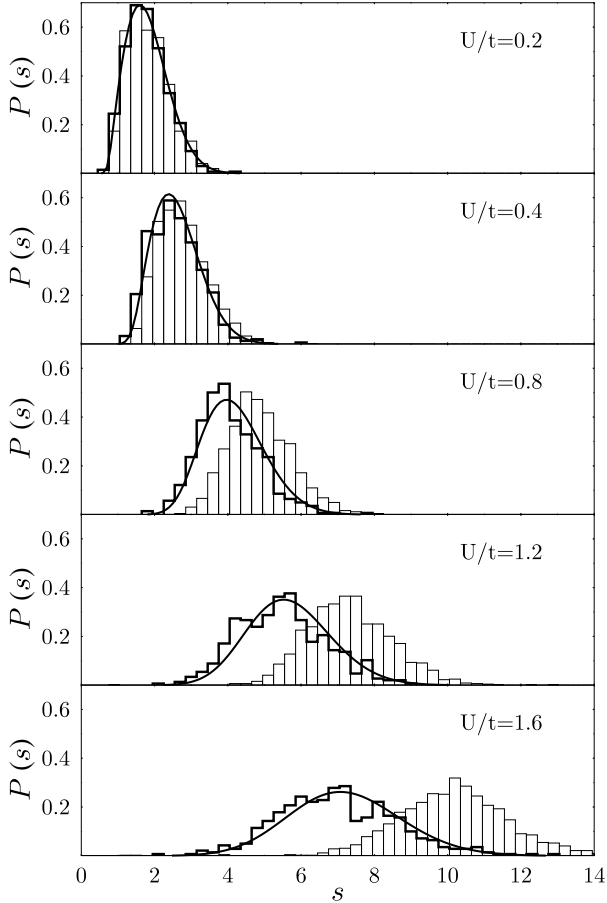,width=3.85in}} 
\end{picture}   
\caption{Probability density distributions of the spacing between the levels
across $\epsilon_f$ (histograms) and of $D_2(16)$ (broken lines). The solid
curves correspond to the estimate ({\protect\ref{gaussps}}).}
\label{deefgraph}
\end{figure}

In Fig. \ref{deefgraph} we present the distribution for the spacing $\Delta
\varepsilon $ between the HF levels across the Fermi energy together with
the distribution of $D_{2}(16)$ calculated according to its definition (\ref
{D2def}) and using the HF many body ground state energies for 15,16 and 17
electrons. We also compare them to the analytic estimate (\ref{gaussps}).
For its evaluation we used $\overline{\Delta }$ that interpolates between
the values found from the Wigner functions fitting $P(s)$ below and above $%
\varepsilon _{f}$ and the numerical results for $\langle J\rangle $ and $%
\sigma (J)$. The two distributions evolve from shifted Wigner functions at
small values of $U/t$ to an approximate Gaussian distributions as the
interaction strength is increased. The crossover occurs when $r_{s}\sim
U/t\sim 1$. Around this point the fluctuations of the Coulomb gap $\sigma
(2J)$ are comparable to the width of the Wigner function describing $P(s)$
at the vicinity of the Fermi energy (see Fig. \ref{vargraph}). Consequently
the latter is smeared into a new more symmetric and broader distribution. It
is evident from Fig. \ref{deefgraph} that Koopmans' theorem breaks down for
strong interactions i.e. $P(\Delta \varepsilon )\neq P(D_{2})$. However it
seems that the two distributions may be viewed as shifted versions of each
other having similar shapes but somewhat different widths (see Fig. \ref
{vargraph}). The shift of $P(D_{2})$ towards lower values is expected
and is due to the change of the occupied levels in response to the
additional electron. This rearrangements, which is neglected by Koopmans'
theorem, tends to lower the electrostatic charging energy. For reasons that
are not clear to us our analytic estimate for $P(\Delta \varepsilon )$ fits
rather well $P(D_{2})$.

\begin{figure}
  \narrowtext 
\setlength{\unitlength}{1in}
\begin{picture}(3.25,2.5)(0,-1)  
\put(-0.4,-1.2){\psfig{figure=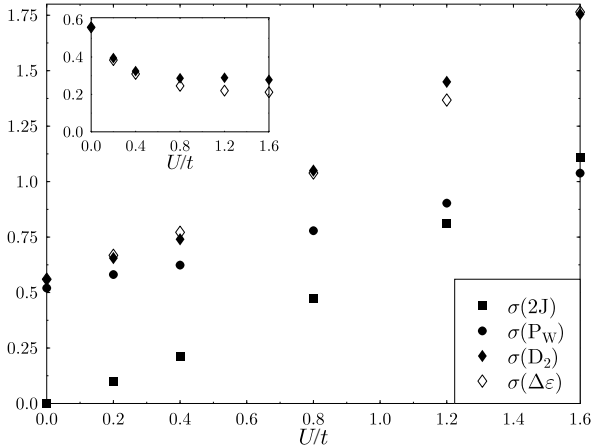,width=4in}} 
\end{picture}     
\caption{Same as Fig. {\protect\ref{vargraph}} but for a dot filled with 
10-12 electrons and disorder strength $W=1.6\, t$.}
\label{vargraph1}
\end{figure}

We repeated the numerical calculations for the same dot but with stronger
disorder $W/2=1.6\, t$ and fewer (10-12) electrons. All of the effects
reported above have been observed for this case as well. The deviation of $%
P(s)$ above or below $\varepsilon_f$ from the Wigner function were more
pronounced. The fluctuations of $J$ also increased and we found $%
\sigma(J)\simeq 0.25\, U/t+0.06(U/t)^2$ with enhanced asymmetry in the shape
of the distribution for strong interactions. The discrepancy between $%
P(\Delta\varepsilon)$ and $P(D_2)$ at large $r_s$ persisted although the
shapes of the two distributions and particularly their width were closer for
this dot then for the one described above. Fig. \ref{vargraph1} summarizes
the dependence of the fluctuations of the different quantities on the
strength of the interaction.

\begin{figure}
  \narrowtext 
\setlength{\unitlength}{1in}
\begin{picture}(3.2,4.8)(0,-1)  
\put(-0.25,-0.8){\psfig{figure=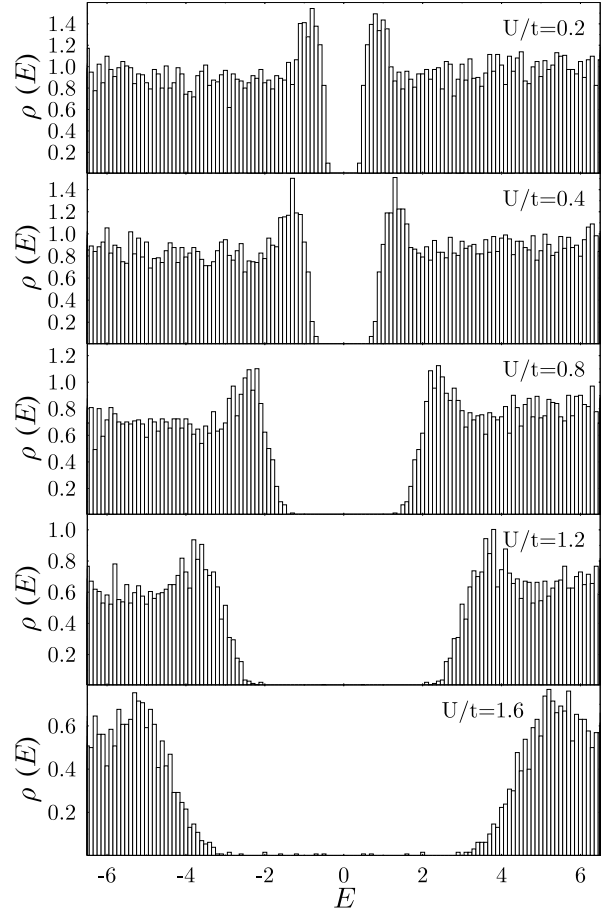,width=3.85in}} 
\end{picture}   
\caption{The density of states near $\epsilon_f$. The results are for a dot
filled with 15-17 electrons and disorder strength $W=1.2\, t$.}
\label{dosfig}
\end{figure}

Our calculations were done for a fixed number of particles. In order to
facilitate comparison with a fixed chemical potential ensemble the Fermi
energy of each spectrum in the ensemble was shifted to zero. The resulting
density of states is plotted in Fig \ref{dosfig}.

\end{multicols}
\renewcommand{\theequation}{A\arabic{equation}}
\widetext

\appendix

\section{The Two Level Hartree-Fock Problem}

In subsection \ref{valid} we outlined the way in which the density matrix $%
\rho$ changes under a variation $\delta h$ of the random part of the HF
Hamiltonian. We argued that while in the case of two close levels that are
both occupied or empty the term $tr(V^A\delta\rho)$ does not alter the small 
$s$ behavior of $P(s)$ it must be included in a self-consistent manner for a
couple of close level residing on both sides of $\varepsilon_f$. In this
appendix we will concentrate on the latter scenario assuming $V\ll\Delta$.

Let us denote by $\psi _{i}$ the eigenfunctions of $h^{(HF)}$ and by $\phi
_{i}$ the eigenfunctions of $h^{(HF)}+\delta h^{(HF)}$. Then in the basis of 
$\psi _{1}$ and $\psi _{2}$ the HF problem for two close levels on both
sides of $\varepsilon _{f}$ is 
\begin{equation}
\begin{array}{c}
\left( \varepsilon_{1}^{(0)}+\delta h_{11}+\sum\limits_{I,J=1,2}\widetilde{V}
_{1I1J}\delta \rho _{JI}\right) a+\left( \delta h_{12}+\sum\limits_{I,J=1,2} 
\widetilde{V}_{1I2J}\delta \rho _{JI}\right) b=\varepsilon a\;,
\label{twlvl} \\ 
\left( \delta h_{12}^{*}+\sum\limits_{I,J=1,2}\widetilde{V}_{2I1J}\delta
\rho _{JI}\right) a+\left( \varepsilon_{2}^{(0)}+\delta
h_{22}+\sum\limits_{I,J=1,2}\widetilde{V}_{2I2J}\delta \rho _{JI}\right)
b=\varepsilon b\;,
\end{array}
\end{equation}
with $\delta \rho _{JI}$ the $2\times 2$ matrix 
\begin{equation}
\delta \rho =\phi _{1}\phi _{1}^{+}-\psi _{1}\psi _{1}^{+}=\left( 
\begin{array}{cc}
|a|^{2}-1 & ab^{*} \\ 
a^{*}b & |b|^{2}
\end{array}
\right) \;,  \label{fneqs}
\end{equation}
and the normalization condition $|a|^{2}+|b|^{2}=1$. The relation between
the matrix $\widetilde{V}_{IJKL}$, $I=1,2$, etc., and the original $%
V_{\alpha \beta \gamma \delta }$ will be discussed below. We will also need
the part of the HF energy the minimization of which, with respect to $\phi
_{1}$, gives (\ref{twlvl}) 
\begin{equation}
E_{HF}[\phi _{1}]=tr\left\{ \left[ h+tr(\widetilde{V}\rho )+\delta h\right]
\phi _{1}\phi _{1}^{+}+\frac{1}{2}tr\left[ \widetilde{V}(\phi _{1}\phi
_{1}^{+}-2\psi _{1}\psi _{1}^{+})\right] \phi _{1}\phi _{1}^{+}\right\} \;.
\label{hfen}
\end{equation}

{\bf Transformation to an effective magnetic problem.} The solution of the
set of algebraic equations (\ref{twlvl}) is particularly simple if one uses
the expansion in terms of Pauli matrices, 
\begin{equation}
\phi_{1}\phi_{1}^{+} = \frac{1}{2}(1+{\bf m}\cdot \mbox {\boldmath $\sigma$}%
)\;\;\;, \;\;\; \psi_{1}\psi_{1}^{+} = \frac{1}{2}(1+{\bf m}_{0}\cdot 
\mbox {\boldmath
$\sigma$})\;,  \label{expan}
\end{equation}
with the "pure state" conditions $|{\bf m}|=|{\bf m}_{0}|=1$. Inserting this
into (\ref{hfen}) and transferring to the left all terms which do not depend
on ${\bf m}$ one obtains 
\begin{equation}
{\cal E}={\bf B\cdot m}-\frac{1}{2}{\bf m}\cdot\stackrel{\leftrightarrow}%
{\bf J} \cdot{\bf m} \; .  \label{reden}
\end{equation}
The notation in (\ref{reden}) stands for 
\begin{eqnarray}
{\cal E}&=&E_{HF}-\frac{1}{2}tr\left[h+tr(\widetilde{V}\rho)+\delta h\right]
+\frac{1}{4}tr\left\{tr\left[ \widetilde{V}({\bf m}_{0}\cdot%
\mbox
{\boldmath $\sigma$})\right]\right\} +\frac{1}{8}tr\left[tr(\widetilde{V}
)\right] \;,  \nonumber \\
{\bf B}&=&\frac{1}{2}tr\left\{\mbox {\boldmath $\sigma$}\left[h+tr(%
\widetilde{V} \rho)+\delta h\right]\right\}-\frac{1}{4}tr\left\{tr\left[
\mbox {\boldmath
$\sigma$} \widetilde{V}(\mbox{\bf m}_{0}\cdot \mbox {\boldmath
$\sigma$})\right]\right\}\;,  \nonumber \\
J_{ab}&=&-\frac{1}{4}tr\left[\sigma^{a}tr(\widetilde{V}\sigma ^{b})\right]\
\; ,  \label{jdef}
\end{eqnarray}
\hfill
\begin{picture}(3.375,0)
  \put(0,0){\line(1,0){3.375}}
  \put(0,0){\line(0,-1){0.08}}
\end{picture}

\begin{multicols}{2}
\noindent
and ${\bf m}\cdot\stackrel{\leftrightarrow}{\bf J}\cdot{\bf m}
=\sum_{a,b=1}^{3} J_{ab}m_{a}m_{b}$. Varying ${\cal E}$ with respect to 
${\bf m}$ under the condition $|{\bf m}|=1$ one obtains 
\begin{equation}
{\bf m}\times ({\bf B}-\stackrel{\leftrightarrow}{\bf J}\cdot {\bf m})=0\;.
\label{vfseapen}
\end{equation}
This is a ``magnetic form'' of the equations (\ref{twlvl}). Of course it
could also be obtained by the direct substitution of (\ref{expan}). The
"magnetic field" ${\bf B}$ contains the information about the random part 
$h+\delta h$ of the HF Hamiltonian whereas the matrix $\stackrel
{\leftrightarrow}{\bf J}$ is the projection on the Pauli matrices of the
interaction matrix $\widetilde{V}_{IJKL}$ taken in the subspace of the two
states $\psi_1$ and $\psi_2$. For our general discussion we can diagonalize 
$\stackrel{\leftrightarrow}{\bf J}$. This will reshuffle the components of 
${\bf B}$ which are random anyway.

The distance $\Delta\varepsilon$ between the eigenvalues of (\ref{twlvl}) 
is $\Delta\varepsilon=tr\left[(\phi_{2}\phi_{2}^{+}-\phi_{1}\phi_{1}^{+})
(h+\delta h)^{HF}\right]$ where due to orthogonality $\phi_{2}\phi_{2}^{+}=
(1-{\bf m}\cdot 
\mbox
{\boldmath $\sigma$})/2$. Therefore 
\begin{equation}
\Delta\varepsilon({\bf B,\stackrel{\leftrightarrow}{J}})=2\left| {\bf B}-{\bf 
\stackrel{\leftrightarrow}{J}\cdot m(B,\stackrel{\leftrightarrow}{J})}
\right| \;,  \label{edif}
\end{equation}
with ${\bf m(B,\stackrel{\leftrightarrow}{J})}$ -- the solution of 
(\ref{vfse}).

The equations are especially simple in the case when the Hamiltonian and the
wavefunctions are real. This corresponds to the orthogonal ensemble in the
terminology of RMT. Only the projections on $\sigma _{1}$ and $\sigma _{3}$
remain in this case. The HF energy and the HF equation are respectively 
\begin{equation}
{\cal E}=B_{1}m_{1}+B_{3}m_{3}-\frac{1}{2}(J_{1}m_{1}^{2}+J_{3}m_{3}^{2})\;,
\label{hfener}
\end{equation}
\begin{equation}
m_{1}B_{3}-m_{3}B_{1}+(J_{1}-J_{3})m_{1}m_{3}=0\;,  \label{m1m3eq}
\end{equation}
with the constraint $m_{1}^{2}+m_{3}^{2}=1$. In the above $J_{1}$ and $J_{3}$
are the eigenvalues of $\stackrel{\leftrightarrow }{\bf J}$. The solution of
(\ref{m1m3eq}) are roots of a fourth order equation. Real roots are the
extrema of the curve defined by the intersection of the surface (\ref{hfener}
) and the constraint cylinder $m_{1}^{2}+m_{3}^{2}=1$. Only one real root
must be selected -- that which minimizes ${\cal E}$. 

Let us examine the energy spacing $\Delta\varepsilon$ 
\begin{equation}
\Delta\varepsilon=2\sqrt{(B_{1}-J_{1}m_{1})^{2}+(B_{3}-J_{3}m_{3})^{2}}
\end{equation}
and determine when and how level crossings occur. In the absence of the
interaction ($J_{1}=J_{3}=0$) $\Delta\varepsilon=0$ at $B_{1}=B_{3}=0$, 
reproducing the standard result. In the presence of the interaction the 
level crossing $\Delta\varepsilon=0$ occurs when 
\begin{equation}
m_{1}=B_{1}/J_{1}\;\;\;,\;\;\;m_{3}=B_{3}/J_{3}\;.  \label{crocon}
\end{equation}
Due to the normalization constraint this can be fulfilled only on an ellipse 
\begin{equation}
\frac{B_{1}^{2}}{J_{1}^{2}}+\frac{B_{3}^{2}}{J_{3}^{2}}=1\;.  \label{ellips}
\end{equation}
It is therefore sufficient to examine the minimum energy solution on the
ellipse. We note that the conditions (\ref{crocon}) are consistent with
equation (\ref{m1m3eq}), i.e. level crossings must occur at least for one of
the solutions. However there is no guarantee that this is also the minimum
energy solution. In fact a detailed analysis reveals that the occurrence of
level crossing for the minimum energy solution depends on the signs of the
eigenvalues $J_{1}$ and $J_{3}$ as follows: ({\it i}) when both signs are
positive there are four real solutions (two minima and two maxima) on the
ellipse (\ref{ellips}) provided $J_{3}>2J_{1}$ or $J_{1}>2J_{3}$.
Otherwise there are only two real solutions (one maximum and one minimum).
The solution with $\Delta\varepsilon=0$ is always the (highest) maximum. 
({\it ii}) when one of the eigenvalues is negative and another positive 
there are four real solutions on the ellipse (two minima and two maxima). 
The solution with $\Delta\varepsilon=0$ is either a metastable minimum or 
a maximum. ({\it iii}) only when both eigenvalues are negative the solution 
with $\Delta\varepsilon=0$ coincides with the absolute minimum of the HF 
energy.

In the first two cases the minimum energy solution always has a non-zero
level spacing. Consequently for such $J_{1}$ and $J_{3}$ the two HF levels 
{\em never cross} and one expects a {\em gap} in the probability
distribution of $\Delta\varepsilon$. In the last case of two positive 
eigenvalues the level spacing distribution in the vicinity of 
$\Delta\varepsilon=0$ will depend on the density of 
$\Delta\varepsilon=const$ lines close to the ellipse (\ref{ellips}).

The interaction $\widetilde{V}$ which determines $J_{1}$ and $J_{3}$ is just 
$V^{A}$ when the two levels $\varepsilon_{1}$ and $\varepsilon_{2}$ are not
degenerate. In this case we find from (\ref{jdef}) 
$J_{ab}=[(V_{1212}-V_{1221})/2]\delta_{ab}$. Thus for a positive (negative)
definite \cite{Lieb}, i.e., repulsive (attractive) interaction 
$J_{1}=J_{3}>0 $ ($J_{1}=J_{3}<0$). When the levels are degenerate one must
extend the two levels treatment to include all degenerate wavefunctions.
However, if the degeneracy is due to spin and the wavefunctions are
separable spin--space products, one can define an equivalent two level
problem for the orbital parts with the interaction $\widetilde{V}=V^{A}+V$.
For this case $J_{1}=(V_{1212}-2V_{1221}-V_{1122})/2$ and 
$J_{3}=[2V_{1212}-V_{1221}-(V_{1111}+V_{2222})]/2$. For such $\widetilde{V}$
the signs of $J_{1}$ and $J_{3}$ are not uniquely related to the nature of
the interaction. In particular they may both be positive even for a
repulsive interaction. However, if this happens, we expect to find, due to a
general argument \cite{Lieb}, another HF solution (e.g. lacking spin
degeneracy) with lower total energy for which the levels do not cross.

The case $J_{1}=J_{3}=J$ is particularly simple since the ellipse
degenerates into a circle and one obtains 
\begin{equation}
\Delta\varepsilon(B,J)=2|B+J|\;,  \label{desimp}
\end{equation}
where $B=\sqrt{B_{1}^{2}+B_{3}^{2}}$. For positive $J$ this result leads,
after choosing for simplicity $P(B)$ of a Gaussian form, to the shifted 
Wigner distribution, Eq. (\ref{shiftwig}). For negative $J$ (attractive 
interaction) we find
\end{multicols}
\renewcommand{\theequation}{A\arabic{equation}}
\widetext

\noindent
\setlength{\unitlength}{1in}
\begin{picture}(3.375,0)
  \put(0,0){\line(1,0){3.375}}
  \put(3.375,0){\line(0,1){0.08}}
\end{picture}
\begin{equation}  \label{psjatt}
P(s;J)= \left\{ 
\begin{array}{cc}
\frac{\pi}{2\Delta^2}\left[(2|J|-s)e^{-\frac{\pi}{4}\left(\frac{2|J|-s}
{\Delta}\right)^2}+(2|J|+s)e^{-\frac{\pi}{4}\left(\frac{2|J|+s}{\Delta}
\right)^2}\right] & s<2|J| \\ 
\frac{\pi}{2\Delta^2}(2|J|+s)e^{-\frac{\pi}{4}\left(\frac{2|J|+s}{\Delta}
\right)^2} \;\; & s\ge 2|J|
\end{array}
\right. \; .
\end{equation}
\hfill
\begin{picture}(3.375,0)
  \put(0,0){\line(1,0){3.375}}
  \put(0,0){\line(0,-1){0.08}}
\end{picture}

\begin{multicols}{2}
\noindent 
This distribution is a sum of two, mirror reflected and shifted Wigner 
distributions resulting from the two roots of $s=\Delta\varepsilon(B,J)$. 
The most important feature to note in (\ref{psjatt}) is the non-zero 
probability $P(s=0)$. This is a consequence of the linear dependence of 
$\Delta\varepsilon(B,J)$ at $B=|J|$. 

\centerline{\bf Acknowledgment} We benefited from the valuable help of M.
Milgrom. We greatfully acknowledge useful discussions with Y. Levinson, O.
Bohigas, E. Domany, A. Finkelstein, Y. Gefen, A. Kamenev, D. Khmelnitskii
and A. Mirlin.

\end{multicols}

\end{document}